\documentclass{nature}
\usepackage{graphicx}
\usepackage{amsmath}

\author{David A. Valverde-Ch\'{a}vez, Carlito Ponseca Jr.$^2$, Constantinos Stoumpos$^3$, Arkady Yartsev$^2$, \\Mercouri G. Kanatzidis$^3$, Villy Sundstr\"{o}m$^2$  \& David G. Cooke$^1$}

\begin{document}

\title{Intrinsic femtosecond charge generation dynamics in a single crystal organometal halide perovskite}
\maketitle
\begin{affiliations}
 \item Department of Physics, McGill University, Montreal, QC, Canada H3A 2T8
 \item Division of Chemical Physics, Lund University, Box 124, 221 00 Lund, Sweden
 \item Department of Chemistry, Northwestern University, Evanston, Illinois 60208, United States
\end{affiliations}


\begin{abstract}
\section*{Abstract:}
Hybrid metal-organic perovskite solar cells have astounded the solar cell community with their rapid rise in efficiency over the past three years. Despite this success, the basic processes governing the photogeneration of free charges, particularly their dynamics and efficiency, remain unknown. Here we use ultrabroadband pulses of THz frequency light to see the intrinsic photophysical properties of single crystal lead halide perovskite just femtoseconds after a photon is first absorbed. Our spectra reveal the dynamics and efficiencies of free charge creation, the remarkable ease in which they move through the lattice and the complicated interplay between free and bound charges in these materials.
\end{abstract}


The stellar rise of hybrid metal halide perovskites displaying impressive photovoltaic and water photolysis efficiencies \cite{LiuNature2013, LuoScience2014}, combined with a composition of earth abundant elements, has lead to a flurry of spectroscopic work probing their underlying photophysics. Only very recently have large single crystals become available, allowing a glimpse into their intrinsic optical and electronic properties \cite{StoumposInorChem2013, DongScience2015, ShiScience2015, NieScience2015}. A key spectral region for these materials is 1 - 100 meV, encompassing proposed exciton binding energies that determine charge photogeneration efficiency\cite{DInnocenzoNatComm2014, SavenijeJPhysChem2014}, carrier scattering rates governing charge transport, and lattice vibrations whose role in screening the exciton remains controversial \cite{PoglitschJChemPhys1987, LinNatPhoton2014}. In this work, we apply time-resolved multi-terahertz spectroscopy to probe photoexcitations over this entire spectral range in single crystal CH$_3$NH$_3$PbI$_3$ with sub-100 fs resolution. We observe ultrafast dynamic screening of lattice phonons by free charges with remarkably high charge mobilities of 500-800 cm$^2$/Vs. Optical pumping at the band edge, where exciton dissociation occurs purely through coupling to the thermal bath, reveals an exciton binding energy of \textbf{$49\pm3$} meV.  At high photon flux, a dynamic exciton Mott transition is observed as mobile charges screen the insulating exciton gas leading to dissociation into a conducting plasma on a sub-1 ps time scale. Our measurements pave the way for a new understanding of fundamental excitations in this material, providing a direction toward optimization of this promising class of solar cells.

The overall power conversion efficiency of organo-metal halide perovskite solar cells has climbed to 20.1\% in a span of only three years \cite{ZhouScience2014, GratzelNatMater2014}, the fastest rate of increase compared to all other solar cell technologies. This fast pace in device development has been matched by an effort to understand the intrinsic charge generation, recombination and transport properties of these materials. Among the key questions remaining are the exciton binding energy \cite{DInnocenzoNatComm2014, SavenijeJPhysChem2014, LinNatPhoton2014}, the exciton/free charge generation dynamics \cite{DInnocenzoNatComm2014, PonsecaJACS2014}, the energetics of lattice excitations and their role in exciton dissociation \cite{JuarezPerezJPCL2014, LinNatPhoton2014}. Time-resolved terahertz spectroscopy (TRTS) is a powerful technique to probe these low energy excitations immediately after photoexcitation on femtosecond time scales \cite{CookePRL2012}. Previous THz studies on the trihalide and mixed halide perovskites have focused on frequency-averaged THz absorption, probing energies below 10 meV \cite{WehrenfennigAdvMater2014, PonsecaJACS2014, SavenijeJPhysChem2014}. These measurements on thin film samples have placed initial lower limits on the mobility, shown suppressed bi-molecular recombination, and indicated the role of the exciton binding energy in the generation of free charges on a $\sim 2$ ps time scale \cite{PonsecaJACS2014}. Here we perform complete energy and time-resolved ultra-broadband THz spectroscopy in the 8-100 meV spectral range with 40 fs temporal resolution after photoexcitation on a single crystal of the hybrid metal halide perovskite CH$_3$NH$_3$PbI$_3$. Photoexcitation at the band edge probes the kinetics of mobile charge generation via dissociation of excitons without the influence of injected excess kinetic energy. Our data reveal the generation dynamics and quantum yields of free charges via thermal dissociation governed by an exciton binding energy of $49\pm3$ meV. We further observe remarkable carrier mobilities that reach as high as 800 cm$^2$/Vs on ultra-short time scales, and the dynamic screening of the exciton at high photon flux.

\begin{figure*}
\includegraphics[width=11.5cm, angle=90]{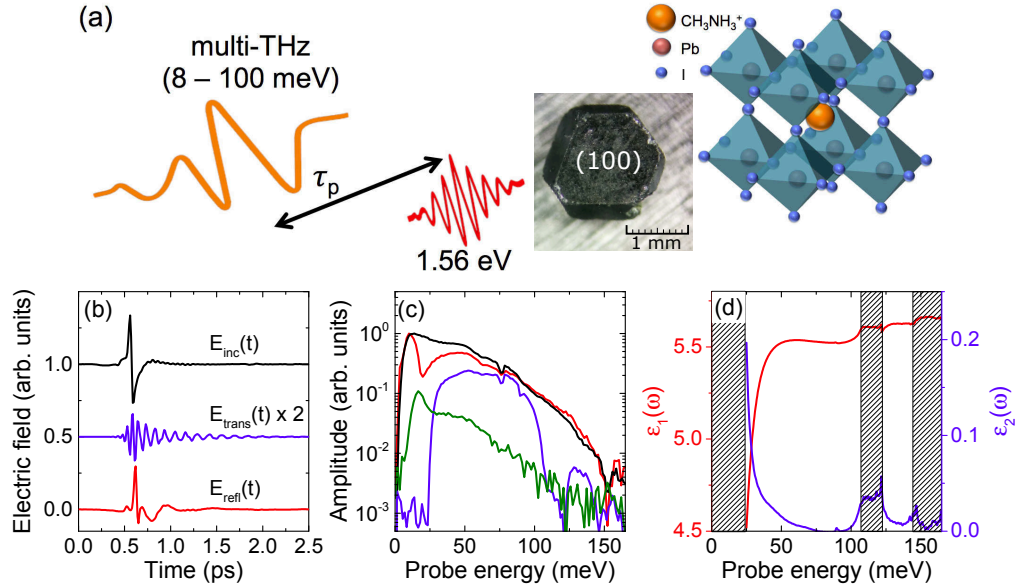}%
\caption{\label{Fig1} \textbf{Multi-THz spectroscopy of CH$_3$NH$_3$PbI$_3$} (a) A schematic of the time-resolved multi-THz spectroscopy experiment is given, where a 40 fs, 1.56 eV pump pulse photoexcites the CH$_3$NH$_3$PbI$_3$ crystal at normal incidence, colinearly with  a multi-THz probe pulse with Fourier spectral components spanning 1 - 30 THz (4 - 125 meV). The sample is a single crystal CH$_3$NH$_3$PbI$_3$ approximately 1.5 x 1.5 mm in size with the (100) facet as the probed surface. (b) In the absence of a photoexcitation, the incident (black), transmitted (blue) and reflected (red) multi-THz pulses after interaction with the crystal are shown with (c) corresponding Fourier amplitude spectra. The amplitude of the pump-induced differential reflected field, $\vert\Delta E(\omega, 600 fs) \vert$, is also shown in green. (d) The dark state dielectric function probing the (100) plane in transmission, with areas of no information indicated by the hatched regions. }
\end{figure*}

Fig 1(a) shows the schematic of the optical pump - THz probe measurement of the large, mm$^3$ sized single crystal of CH$_3$NH$_3$PbI$_3$, probing the (100) facet of the crystal with image provided. The unit cell of the cubic phase is also shown, with the average position of the CH$_3$NH$_3^+$ cation indicated by the orange ball in the center of the unit cell. In these experiments, an optical pulse photoexcites the sample and the induced charge degrees of freedom are then probed by transmission or reflection of a phase-stable, single cycle electromagnetic transient with frequency components in the THz range, called a THz pulse. The incident THz pulse (E$_\text{inc}$), is shown in Fig. 1(b) along with the transmitted (E$_\text{trans}$) and reflected (E$_\text{refl}$) pulses. The ultra-broadband THz pulse E$_\text{inc}$(t) contains spectral components from 1 - 30 THz (4-125 meV or 33-1000 cm$^{-1}$), as shown in the corresponding Fourier amplitude spectrum in Fig 1(c). While the reflected pulse retains all incident Fourier components with noticeable dispersion, E$_\text{trans}$(t) is strongly chirped with no spectral components below 25 meV due to absorption and reflection arising from an infrared active phonon and dielectric relaxation of the CH$_3$NH$_3^+$ cation \cite{EvenJPhysChemC2014}. The change of amplitude and phase of the THz pulse after transmission provides the complex dielectric function $\tilde{\epsilon}(\omega)=\epsilon_1(\omega)+i\epsilon_2(\omega)$ via inversion of the Fresnel equations, and is shown in Fig. 1(d). The spectra is dominated by an Pb-I bending mode at  $\sim8$ meV previously observed in Raman and THz spectroscopy \cite{QuartiJPhysChemLett2014, WehrenfennigEnEnvSci2014}. In addition, a broad GHz dielectric contribution has also been identified from microwave measurements due to the relaxation of the CH$_3$NH$_3^{+}$ cation \cite{PoglitschJChemPhys1987}. The CH$_3$NH$_3^{+}$ group is free to rotate and undergo librations in the room temperature tetragonal phase giving a Debye contribution to the dielectric function with a picosecond relaxation time. As a result of these two lattice responses, the optical transmission is strongly suppressed for energies below 25 meV. The real dielectric function, $\epsilon_1(\omega)$, responsible for renormalizing the Coulomb interaction of injected electrons and holes and therefore the exciton binding energy E$_\text{B}$ \cite{GreenNatPhoton2014, SabaNatComm2014}, is nearly constant at 5.5 for 40 - 100 meV range. Below 40 meV near resonant with the exciton binding energy, however, $\epsilon_1(\omega)$ is quite dispersive. The dynamics of exciton-phonon coupling should therefore be considered when discussing the appropriate dielectric function \cite{EvenJPhysChemC2014}.

We now turn to the time-resolved THz spectra after photoexcitation. The reference multi-THz waveform in the absence of optical pumping, E$_\text{ref}$(t), is shown in Fig. 1(b) and Fig. 2(a) after normal incidence reflection from the (100) crystal facet, showing slight reshaping due to the dispersion of the crystal. The power spectrum of this pulse, shown in Fig. 1(c), contains frequency components spanning the entire 30 THz bandwidth. The pump-induced change in the reflected THz electric field $\Delta$E(t,$\tau_p) = $E$_\text{pump}(t,\tau_p)$-E$_\text{ref}(t)$ at a given pump-probe delay time $\tau_p$ is measured at variable delays producing the two-dimensional time-domain data set shown in Fig. 2(b). The Fourier amplitude of the modulated field in quasi-equilibrium is given in Fig. 1(c), showing spectral information is obtained between 8 - 80 meV. The significant phase shift compared to the reference pulse indicates a strong change in the dispersion after charges are photo-injected into the crystal. This is a signature of injected mobile charges screening the lattice from the applied field, within the pump pulse penetration depth of 1 $\mu$m in the crystal \cite{GreenNatPhoton2014}. The amplitude of the $\Delta$E(t,$\tau_p)$ response signifies the onset of photoconductivity in the sample, and reaches its maximum within 400 fs after photoexcitation, seen in the constant $\tau_p$ waveforms in Fig. 2(c). We note that this is 10 times the temporal resolution of our technique and thus mobile carriers are not generated instantaneously but rather through exciton dissociation \cite{SavenijeJPhysChem2014}.

\begin{figure}
\includegraphics[width=8.9cm]{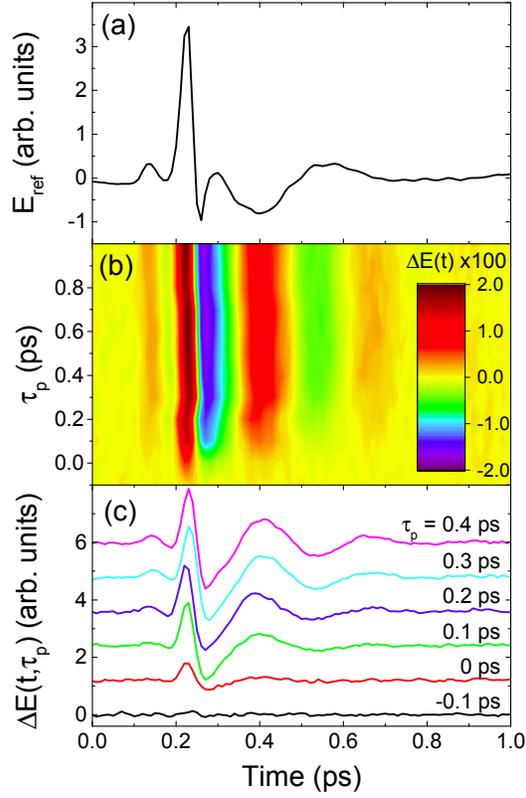}%
\caption{\label{Fig2} \textbf{THz transients and pump-induced differential waveforms} (a) The reflected THz transient from the unexcited CH$_3$NH$_3$PbI$_3$ crystal (100) facet and (b) the two-dimensional $\Delta$E(t,$\tau_p$) = E$_\text{pump}$(t,$\tau_p$)-E$_\text{ref}$(t) data set for pump-probe delay times $\tau_p < 1$ ps. (c) Selected cuts in the map along constant $\tau_p$ are shown, representing the buildup of photoconductivity in the crystal.}
\end{figure}

The corresponding spectral amplitude and phase of the pump-induced change in THz reflectance, defined as $\Delta\tilde{r}(\omega,\tau_p) = \Delta\tilde{E}(\omega,\tau_p)/\tilde{E}_{ref}(\omega)$, are shown in Fig. 3(a) and (b), respectively for pump-probe times from 0 - 400 fs at a pump fluence of 80 $\mu$J/cm$^2$. The data can be understood by the analytic formula \cite{CookePRL2012}:
\begin{equation}
\Delta\tilde{r}(\omega,\tau_p)=-\biggl(\frac{1+\tilde{r}_0}{\tilde{r}_0}\biggr)\frac{Z_0d\tilde{\sigma}(\omega,\tau_p)}{1+\tilde{n}(\omega)+Z_0d\tilde{\sigma}(\omega,\tau_p)}
\end{equation}
where $\tilde{\sigma}=\sigma_1+i\sigma_2$ is the complex ac photoconductivity, $\tilde{n} = n+i\kappa$ is the complex crystal index of refraction and $\tilde{r}_{0}$ is the complex crystal reflection coefficient given by $\tilde{r}_0 = (1-\tilde{n})/(1+\tilde{n})$. In the absence of photoconductivity, $\Delta\tilde{r}(\omega) = 0$ as well as in the limit for a highly reflective crystal with $n>>1$ where $r_0\rightarrow -1$. Fig. 3(c) shows the complex index of refraction of the crystal used to describe $\tilde{r}_0$, composed of a picosecond Debye dielectric relaxation and Lorentzian response modelling the 8 meV infrared active phonon in agreement with Fig. 1(d). The high energy index of 2.35 is chosen to agree with our transmission data of Fig. 1(b) ($\epsilon_{\infty}\approx 2.35^2 = 5.52$) and low energy limit from literature $\epsilon_s\approx 30$ \cite{PoglitschJChemPhys1987}. The corresponding reflectivity amplitude $\vert r_0\vert$ is shown in Fig. 3(c) exhibiting a sharp drop to a minimum just below 20 meV. Thus the screening of the lattice by injected mobile charge carriers is responsible for the large $\Delta r$ response observed in Fig. 3(a), and by Kramers-Kronig the corresponding phase change in Fig. 3(b). By inversion of Eq. 1, the complete $\tilde{\sigma}(\omega,\tau_p)$ spectra can be extracted in temporal slices following photoexcitation, revealing the underlying ultrafast charge transport characteristics.

An example of the $\tilde{\sigma}(\omega,\tau_p)$ spectrum recorded at $\tau_p =1$ ps after excitation is shown in Fig. 3(d) by inversion of Eqn. 1. The data is very well described by a simple Drude model $\tilde{\sigma}(\omega)=\frac{n_De^2\tau}{m^*}\frac{1}{1-i\omega\tau}$, where $n_D$ is the mobile charge density and $\tau$ is the momentum scattering time determining the charge carrier mobility through the relation $\mu=e\tau/m^*$ with $m^* = 0.2 m_e$ being the charge carrier effective mass \cite{UmariSciRep2014}. The Drude response confirms the band nature of transport in the single crystal CH$_3$NH$_3$PbI$_3$ as opposed to charges in a disordered film where hopping conductivity occurs. The scattering time $\tau$ can be estimated simply by the crossing point of $\sigma_1$ and $\sigma_2$, occurring at an energy $E_c=h/(2\pi\tau) \approx 10$ meV. This provides an estimate of $\tau\sim$70 fs corresponding to a remarkably high $\mu \sim 620$ cm$^2$/Vs for a solution processed semiconductor. This measurement represents the highest mobility observed in these materials to date although recent improvements in grain size and large crystal fabrication have seen $\mu$ on the order of 150 cm$^2$/Vs.\cite{ShiScience2015, DongScience2015}  Moreover, the amplitude of the photoconductivity directly yields the mobile carrier density $n_D\approx3.3\times10^{17}$ cm$^{-3}$, estimated by the Drude dc limit $\sigma_{dc}=n_De\mu$. The mobile charge carrier generation efficiency is given by $n_D/n_{ph} \approx 12 \%$ where $n_{ph}=2.9\times10^{18}$ cm$^{-3}$ is the absorbed photon density given the pump fluence of 80 $\mu$J/cm$^2$ and the 10\% reflection of the pump at 795 nm \cite{LoperJPhysChem2014}. This result is consistent with the expected result of the Saha-Langmuir equation,\cite{DInnocenzoNatComm2014} where a mobile carrier yield of 13\% is expected given a 50 meV exciton binding energy at 293 K and an exciton effective mass $m_{ex} = 0.11$ m$_e$ \cite{UmariSciRep2014}.

\begin{figure*}
\includegraphics[width=12cm, angle=90]{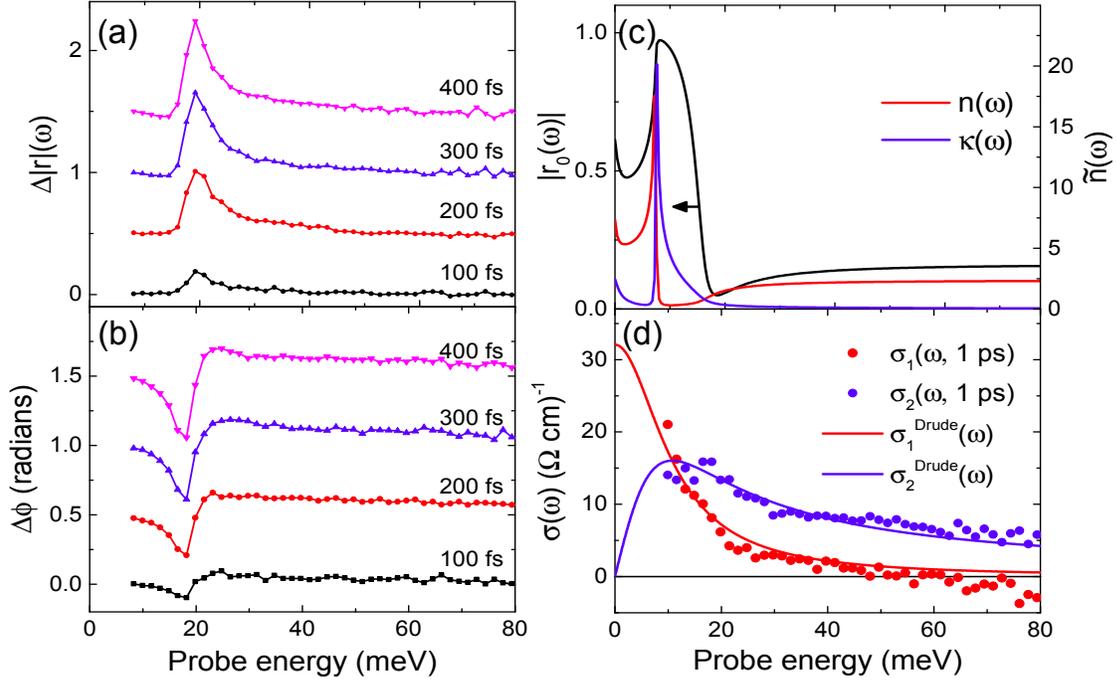}%
\caption{\label{Fig3} \textbf{Free charge carrier screening of IR phonons and ultrafast Drude conduction.} (a) Amplitude and (b) phase of the differential multi-THz reflectance shifted by steps of 0.5 for clarity. (c) The model of the complex index of refraction of the unexcited crystal used to describe the substrate dispersion and extract the complex photoconductivity. (d) The complex conductivity spectrum with Drude spectrum fits recorded at 1 ps following excitation.}
\end{figure*}

\begin{figure*}
\includegraphics[width=12cm, angle = 90]{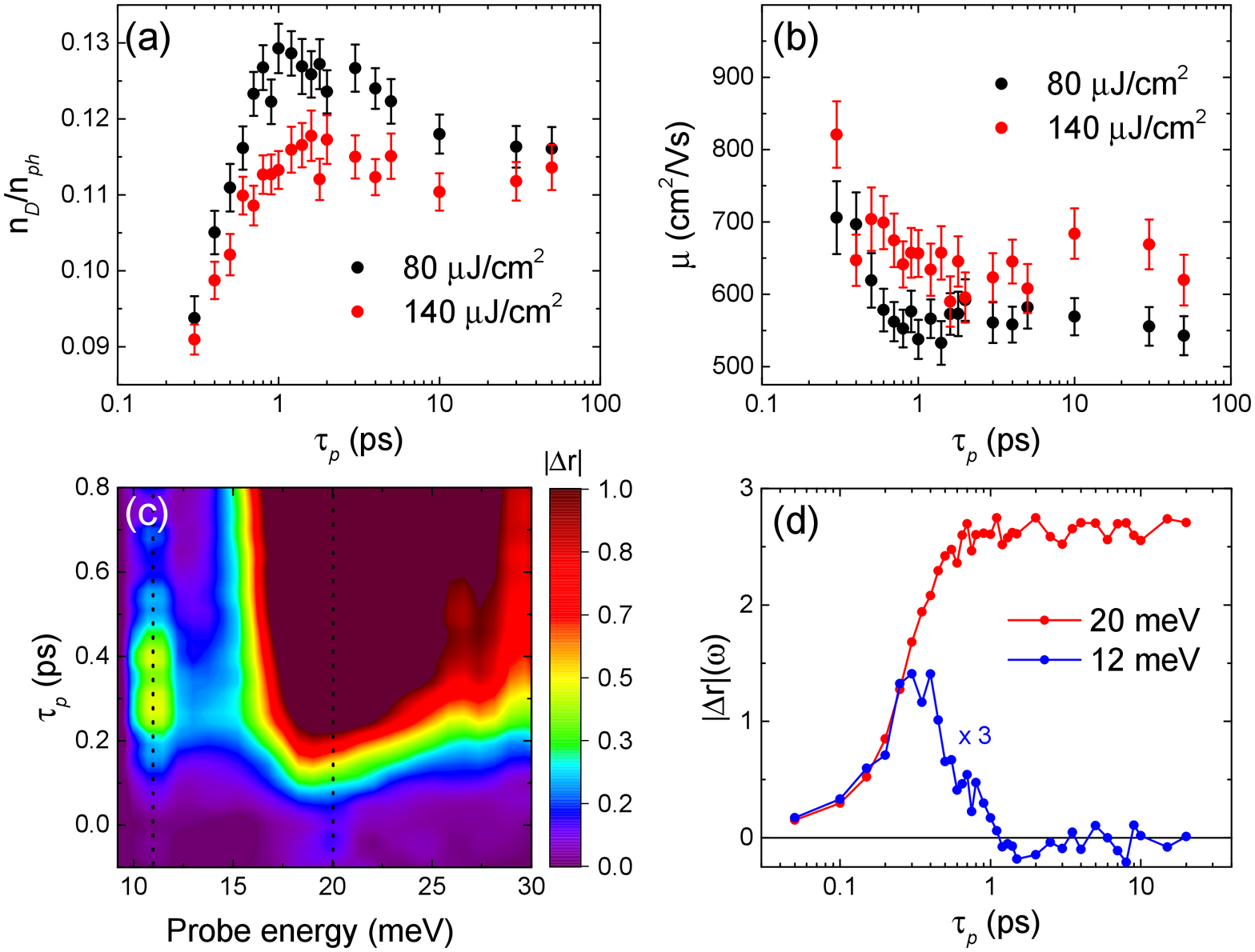}%
\caption{\label{Fig4} \textbf{Free charge generation, charge carrier mobilities and exciton dissociation.} (a) The mobile charge carrier generation efficiency defined as the ratio of the charge density to the absorbed photon density, at pump fluences of 80 and 140 $\mu$J/cm$^2$. (b) Charge carrier mobility derived from the 80 and 140 $\mu$J/cm$^2$ data sets with global complex fits to the Drude model. (c) The two-dimensional data set of the differential reflectance amplitude at a fluence of 810 $\mu$J/cm$^2$, showing the early time region where a low energy transient peak appears at $\approx 12$ meV. Vertical lines indicate cut regions shown in (d) at 12 meV and 20 meV corresponding to exciton and free charges, respectively. The dynamics of exciton dissociation are represented by the decay of the 12 meV peak concomitant with the 1 ps free charge density risetime.}
\end{figure*}

The results of global fitting of the Drude model to the complete two-dimensional time-frequency $\tilde{\sigma}(\omega)$ maps, fitting both real and imaginary components simultaneously, are shown in Fig. 4. The dynamics of the mobile charge generation efficiency for two pump fluences is shown in Fig. 4(a), exhibiting a slight transient enhancement for $\tau_p<10$ ps at the lower fluence. The Saha-Langmuir equation accounts for this behaviour predicting a decrease in efficiency to 10$\%$ for the elevated fluence due to the greater probability of electron-hole binding \cite{DInnocenzoNatComm2014}. The remaining $\sim 90\%$ of excitations, however are bound excitons that are unable to screen the lattice directly due to their charge neutrality. If pump photons excite carriers above the band edge, however, exciton dissociation will be influenced by the initial excess kinetic energy, providing higher mobile charge yields. The goal of this study, however, is to probe the energetics of thermal dissociation governed solely by the exciton binding energy and so we leave this to future work. The mobility is also extracted through the Drude fits as $\mu=e\tau/m^*$ and is found to be relatively fluence independent, as shown in Fig. 4(b). A $\approx 20\%$ decrease is observed in carrier mobility on a sub-picosecond time scale, which we conjecture could be due to the charging of trap states or some charge-mediated phonon scattering mechanism. Despite the slight reduction in the transient mobility, the steady-state mobility of 550 cm$^2$/Vs is orders of magnitude higher than other solution processed semiconductors.

The induced mobile charges, however, should also screen the exciton as discussed in a recent observation of band filling effects \cite{ManserNatPhoton2014}, and so one expects more efficient generation at very high carrier densities. The critical Mott density that defines the threshold for the efficiency of this screening, however, has been estimated at $n_{crit}\approx 1\times10^{18}$ cm$^{-3}$, nearly an order of magnitude higher than densities in Fig. 3. To investigate the effects of charge screening of the exciton, a two-dimensional data set was taken at F$=810\mu$J/cm$^2$ with the $\vert\Delta r(\omega)\vert$ map shown in Fig. 4(c). The response is not well described by a Drude model alone, as it exhibits a transient low energy peak at 12 meV, below the main reflection peak at 20 meV. The dynamics of this excitation are shown by the cut in the map along the $\tau_p$ axis for the two peaks. The onset of the lattice screening occurs on a time scale of 0.5 - 1 ps, larger than the 40 fs temporal resolution of the measurement and indicating the charge generation is driven by exciton dissociation. The 12 meV peak initially tracks the rise time of the free charge 20 meV peak, however decays to the noise level by the time the lattice response is fully screened at $\tau_p=1$ ps. This indicates a coupling between the two excitations and we conclude that the 12 meV peak is due to an intra-excitonic orbital transition given the binding energies for trap states have been recently estimated to be much larger, on the order of 100 - 400 meV from the band edge.\cite{WuJACS2015} We note if excitation into trap states was the dominant photoexcitation at the band edge, one would not expect a Drude conductivity spectrum but rather a conductivity which is activated at a characteristic frequency. Even if the quantum efficiency reaches 4.5\% predicted by the Saha-Langmuir equation at this elevated fluence, $n_{D}\approx 1.3\times 10^{18}$ cm$^{-3}$ which exceeds $n_{crit}$. In this regime, one expects all excitons are dissociated forming an uncorrelated plasma. The disappearance of the 12 meV peak is therefore due to the dynamic screening of the injected exciton population as the free charge density increases beyond the critical Mott density on a sub-picosecond time scale.

In conclusion, we have performed optical pump - multi-THz spectroscopy in the energetically relevant 8 - 100 meV spectral region on a single crystal lead halide perovskite. Our spectra reveal the dynamic screening of infrared active phonons and background Debye relaxation of the CH$_3$NH$_3^+$ cation, due to the photogenerated free charge population. Sub-picosecond conductivity spectra directly reveals free charge densities and remarkably high mobilities of $\sim800$ cm$^2$/Vs on sub-picosecond time scales. Charge generation proceeds via exciton dissociation which is complete after 1 ps, with quantum yields of free carriers governed by a 50 meV binding energy. A 12 meV signature of the exciton is observed, dynamically screened on a time scale comparable to the generation of mobile carriers and at a carrier density consistent with the excitonic Mott transition.

\begin{methods}
\subsection{Sample preparation}

Single-crystals of CH$_3$NH$_3$PbI$_3$ suitable for characterization were grown from a solvent mixture comprising of  aqueous HI (57\% w/w, 5.1 mL) and aqueous H$_3$PO$_2$ (50\% w/w, 1.7 mL). In a typical procedure, a 20 mL scintillation vial was charged with the colorless solvent mixture and heated to boiling (ca. 120$^{\circ}$ C). Addition of solid PbO (670 mg, 3 mmol) and CH$_3$NH$_3$Cl (202 mg, 3 mmol) led initially to the formation of a black precipitate which rapidly dissolved leading to a clear bright yellow solution. The boiling hot solution was then capped with several layers of parafilm tape to ensure that the composition of the vapors was maintained throughout the crystal growth process. Omission of this step leads to the crystallization of CH$_3$NH$_3$PbI$_3$.H$_2$O, instead. On standing, upon reaching ambient temperature a countable number (in the range of 10-100) of small CH$_3$NH$_3$PbI$_3$ crystals begin to form, which act as initial seeds for the subsequent crystal growth from the supersaturated supernatant solution. Well-formed, faceted crystals of rhombic dodecahedral crystal habit were obtained after 2 weeks with crystal sizes ranging from 1-4 mm among several batches, with the size of the crystals being inversely proportional to the number of the initial crystal seeds. The crystals were collected manually by decanting the mother liquor, pressed dry with a soft filtration paper and thoroughly dried under a stream of N$_2$ gas.

\subsection{Characterization}

The 795 nm, 40 fs pulse output of a Ti:sapphire regenerative amplifier is split into three beams. Charge carriers are injected the pump beam resonant with the band edge at 1.56 eV to minimize initial excess energy, thereby suppressing direct exciton dissociation through this channel \cite{KaindlNature2003}. The penetration depth of the pump pulse is taken to be 1 $\mu$m determined by thin film absorption, and it is assumed this is the same for a single crystal. Phase stable THz pulses are generated via a two-colour laser plasma in dry air and the electric field is directly detected after reflection off the crystal facet by an air-biased coherent detection scheme with 30 THz bandwidth \cite{CookePRL2012}. Both E$_{ref}$(t) and $\Delta$E(t,$\tau_p$) are measured simultaneously through a double modulation scheme, required to minimize the influence of systematic timing variations. All measurements were performed under dry air environment and no evidence of oxidation was observed during the experiments.
The background complex refractive index of the crystal $\tilde{n}(\omega)=\sqrt{\tilde{\epsilon}(\omega)}$ and the dielectric function $\tilde{\epsilon}(\omega)$ is given by the Lorentzian phonon and Debye components
\begin{equation}
\tilde{\epsilon}(\omega)=\epsilon_\infty+\frac{\omega_p^2}{(\omega_0^2-\omega^2)-i\omega\gamma}+\frac{\epsilon_s}{1-i\omega\tau_D})
\end{equation}
The high frequency dielectric function $\epsilon_\infty=5.5$ determined from transmission measurements in Fig. 1(b) and the Pb-I bending mode $\omega_0/2\pi = 1.96$ THz and the Debye relaxation time $\tau_D$ = 2.5 ps. The spectra were found to be relatively insensitive to the phonon relaxation rate, set to a reasonable $\gamma=0.1$ THz.
The Saha-Langmuir equation, used to determine the exciton binding energy from the $x=n_D/n_{phot}$ ratio, is defined as \cite{DInnocenzoNatComm2014}:
\begin{equation}
\frac{x^2}{1-x}=\frac{1}{n_{phot}}\bigl(\frac{2\pi m_{ex} k_B T}{h^2}\bigr)^{3/2}e^{-\frac{E_B}{k_B T}}
\end{equation}

\end{methods}

\bibliography{Perovskite}

\begin{thebibliography}{10}
\expandafter\ifx\csname url\endcsname\relax
  \def\url#1{\texttt{#1}}\fi
\expandafter\ifx\csname urlprefix\endcsname\relax\def\urlprefix{URL }\fi
\providecommand{\bibinfo}[2]{#2}
\providecommand{\eprint}[2][]{\url{#2}}

\bibitem{LiuNature2013}
\bibinfo{author}{Liu, M.}, \bibinfo{author}{Johnston, M.~B.} \&
  \bibinfo{author}{Snaith, H.~J.}
\newblock \bibinfo{title}{Efficient planar heterojunction perovskite solar
  cells by vapour deposition}.
\newblock \emph{\bibinfo{journal}{Nature}} \textbf{\bibinfo{volume}{501}},
  \bibinfo{pages}{395--398} (\bibinfo{year}{2013}).

\bibitem{LuoScience2014}
\bibinfo{author}{Luo, J.} \emph{et~al.}
\newblock \bibinfo{title}{Water photolysis at 12.3
  photovoltaics and Earth-abundant catalysts}.
\newblock \emph{\bibinfo{journal}{Science}} \textbf{\bibinfo{volume}{345}},
  \bibinfo{pages}{1593--1596} (\bibinfo{year}{2014}).

\bibitem{StoumposInorChem2013}
\bibinfo{author}{Stoumpos, C.~C.}, \bibinfo{author}{Malliakas, C.~D.} \&
  \bibinfo{author}{Kanatzidis, M.~G.}
\newblock \bibinfo{title}{Semiconducting Tin and Lead Iodide Perovskites with
  Organic Cations: Phase Transitions, High Mobilities, and Near-Infrared
  Photoluminescent Properties}.
\newblock \emph{\bibinfo{journal}{Inorg. Chem.}} \textbf{\bibinfo{volume}{52}},
  \bibinfo{pages}{9019--9038} (\bibinfo{year}{2013}).

\bibitem{DongScience2015}
\bibinfo{author}{Dong, Q.} \emph{et~al.}
\newblock \bibinfo{title}{Electron-hole diffusion lengths $> 175 \mu$m in
  solution-grown CH$_3$NH$_3$PbI$_3$ single crystals}.
\newblock \emph{\bibinfo{journal}{Science}} \textbf{\bibinfo{volume}{347}},
  \bibinfo{pages}{967--970} (\bibinfo{year}{2015}).

\bibitem{ShiScience2015}
\bibinfo{author}{Shi, D.} \emph{et~al.}
\newblock \bibinfo{title}{Low trap-state density and long carrier diffusion in
  organolead trihalide perovskite single crystals}.
\newblock \emph{\bibinfo{journal}{Science}} \textbf{\bibinfo{volume}{347}},
  \bibinfo{pages}{519--522} (\bibinfo{year}{2015}).

\bibitem{NieScience2015}
\bibinfo{author}{Nie, W.} \emph{et~al.}
\newblock \bibinfo{title}{High-efficiency solution-processed perovskite solar
  cells with millimeter-scale grains}.
\newblock \emph{\bibinfo{journal}{Science}} \textbf{\bibinfo{volume}{347}},
  \bibinfo{pages}{522--525} (\bibinfo{year}{2015}).

\bibitem{DInnocenzoNatComm2014}
\bibinfo{author}{D'Innocenzo, V.} \emph{et~al.}
\newblock \bibinfo{title}{Excitons versus free charges in organo-lead
  tri-halide perovskites}.
\newblock \emph{\bibinfo{journal}{Nat. Comm.}} \textbf{\bibinfo{volume}{5}},
  \bibinfo{pages}{1} (\bibinfo{year}{2014}).

\bibitem{SavenijeJPhysChem2014}
\bibinfo{author}{Savenije, T.~J.} \emph{et~al.}
\newblock \bibinfo{title}{Thermally Activated Exciton Dissociation and
  Recombination Control the Carrier Dynamics in Organometal Halide Perovskite}.
\newblock \emph{\bibinfo{journal}{J. Phys. Chem. Lett.}}
  \textbf{\bibinfo{volume}{5}}, \bibinfo{pages}{2189--2194}
  (\bibinfo{year}{2014}).

\bibitem{PoglitschJChemPhys1987}
\bibinfo{author}{Poglitsch, A.} \& \bibinfo{author}{Weber, D.}
\newblock \bibinfo{title}{Dynamic disorder in
  methylammoniumtrihalogenoplumbates (II) observed by millimeter‐wave
  spectroscopy}.
\newblock \emph{\bibinfo{journal}{J. Chem. Phys.}}
  \textbf{\bibinfo{volume}{87}}, \bibinfo{pages}{6373--6378}
  (\bibinfo{year}{1987}).

\bibitem{LinNatPhoton2014}
\bibinfo{author}{Lin, Q.}, \bibinfo{author}{Armin, A.},
  \bibinfo{author}{Nagiri, R. C.~R.}, \bibinfo{author}{Burn, P.~L.} \&
  \bibinfo{author}{Meredith, P.}
\newblock \bibinfo{title}{Electro-optics of perovskite solar cells}.
\newblock \emph{\bibinfo{journal}{Nat. Photon.}} \textbf{\bibinfo{volume}{9}},
  \bibinfo{pages}{106--112} (\bibinfo{year}{2015}).

\bibitem{ZhouScience2014}
\bibinfo{author}{Zhou, H.} \emph{et~al.}
\newblock \bibinfo{title}{Interface engineering of highly efficient perovskite
  solar cells}.
\newblock \emph{\bibinfo{journal}{Science}} \textbf{\bibinfo{volume}{345}},
  \bibinfo{pages}{542--546} (\bibinfo{year}{2014}).

\bibitem{GratzelNatMater2014}
\bibinfo{author}{Gr\"{a}tzel, M.}
\newblock \bibinfo{title}{The light and shade of perovskite solar cells}.
\newblock \emph{\bibinfo{journal}{Nat Mater}} \textbf{\bibinfo{volume}{13}},
  \bibinfo{pages}{838--842} (\bibinfo{year}{2014}).

\bibitem{PonsecaJACS2014}
\bibinfo{author}{Ponseca~Jr., C.~S.} \emph{et~al.}
\newblock \bibinfo{title}{Organometal Halide Perovskite Solar Cell Materials
  Rationalized: Ultrafast Charge Generation, High and Microsecond-Long Balanced
  Mobilities, and Slow Recombination}.
\newblock \emph{\bibinfo{journal}{J. Am. Chem. Soc.}}
  \textbf{\bibinfo{volume}{136}}, \bibinfo{pages}{5189--5192}
  (\bibinfo{year}{2014}).

\bibitem{JuarezPerezJPCL2014}
\bibinfo{author}{Juarez-Perez, E.~J.} \emph{et~al.}
\newblock \bibinfo{title}{Photoinduced Giant Dielectric Constant in Lead Halide
  Perovskite Solar Cells}.
\newblock \emph{\bibinfo{journal}{J. Phys. Chem. Lett.}}
  \textbf{\bibinfo{volume}{5}}, \bibinfo{pages}{2390--2394}
  (\bibinfo{year}{2014}).

\bibitem{CookePRL2012}
\bibinfo{author}{Cooke, D.~G.}, \bibinfo{author}{Krebs, F.~C.} \&
  \bibinfo{author}{Jepsen, P.~U.}
\newblock \bibinfo{title}{Direct Observation of Sub-100~fs Mobile Charge
  Generation in a Polymer-Fullerene Film}.
\newblock \emph{\bibinfo{journal}{Phys. Rev. Lett.}}
  \textbf{\bibinfo{volume}{108}}, \bibinfo{pages}{056603}
  (\bibinfo{year}{2012}).

\bibitem{WehrenfennigAdvMater2014}
\bibinfo{author}{Wehrenfennig, C.}, \bibinfo{author}{Eperon, G.~E.},
  \bibinfo{author}{Johnston, M.~B.}, \bibinfo{author}{Snaith, H.~J.} \&
  \bibinfo{author}{Herz, L.~M.}
\newblock \bibinfo{title}{High Charge Carrier Mobilities and Lifetimes in
  Organolead Trihalide Perovskites}.
\newblock \emph{\bibinfo{journal}{Adv. Mater.}} \textbf{\bibinfo{volume}{26}},
  \bibinfo{pages}{1584--1589} (\bibinfo{year}{2014}).

\bibitem{EvenJPhysChemC2014}
\bibinfo{author}{Even, J.}, \bibinfo{author}{Pedesseau, L.} \&
  \bibinfo{author}{Katan, C.}
\newblock \bibinfo{title}{Analysis of Multivalley and Multibandgap Absorption
  and Enhancement of Free Carriers Related to Exciton Screening in Hybrid
  Perovskites}.
\newblock \emph{\bibinfo{journal}{J. Phys. Chem. C}}
  \textbf{\bibinfo{volume}{118}}, \bibinfo{pages}{11566--11572}
  (\bibinfo{year}{2014}).

\bibitem{QuartiJPhysChemLett2014}
\bibinfo{author}{Quarti, C.} \emph{et~al.}
\newblock \bibinfo{title}{The Raman Spectrum of the CH$_3$NH$_3$PbI$_3$ Hybrid
  Perovskite: Interplay of Theory and Experiment}.
\newblock \emph{\bibinfo{journal}{J. Phys. Chem. Lett.}}
  \textbf{\bibinfo{volume}{5}}, \bibinfo{pages}{279--284}
  (\bibinfo{year}{2013}).

\bibitem{WehrenfennigEnEnvSci2014}
\bibinfo{author}{Wehrenfennig, C.}, \bibinfo{author}{Liu, M.},
  \bibinfo{author}{Snaith, H.~J.}, \bibinfo{author}{Johnston, M.~B.} \&
  \bibinfo{author}{Herz, L.~M.}
\newblock \bibinfo{title}{Charge-carrier dynamics in vapour-deposited films of
  the organolead halide perovskite CH$_3$NH$_3$PbI$_{3-x}$Cl$_x$}.
\newblock \emph{\bibinfo{journal}{Energy Environ. Sci.}}
  \textbf{\bibinfo{volume}{7}}, \bibinfo{pages}{2269--2275}
  (\bibinfo{year}{2014}).

\bibitem{GreenNatPhoton2014}
\bibinfo{author}{Green, M.~A.}, \bibinfo{author}{Ho-Baillie, A.} \&
  \bibinfo{author}{Snaith, H.~J.}
\newblock \bibinfo{title}{The emergence of perovskite solar cells}.
\newblock \emph{\bibinfo{journal}{Nat. Photon.}} \textbf{\bibinfo{volume}{8}},
  \bibinfo{pages}{506--514} (\bibinfo{year}{2014}).

\bibitem{SabaNatComm2014}
\bibinfo{author}{Saba, M.} \emph{et~al.}
\newblock \bibinfo{title}{Correlated electron–hole plasma in organometal
  perovskites}.
\newblock \emph{\bibinfo{journal}{Nat. Commun.}} \textbf{\bibinfo{volume}{5}},
  \bibinfo{pages}{1} (\bibinfo{year}{2014}).

\bibitem{UmariSciRep2014}
\bibinfo{author}{Umari, P.}, \bibinfo{author}{Mosconi, E.} \&
  \bibinfo{author}{De~Angelis, F.}
\newblock \bibinfo{title}{Relativistic GW calculations on CH$_3$NH$_3$PbI$_3$
  and CH$_3$NH$_3$SnI$_3$ Perovskites for Solar Cell Applications}.
\newblock \emph{\bibinfo{journal}{Sci. Rep.}} \textbf{\bibinfo{volume}{4}},
  \bibinfo{pages}{1} (\bibinfo{year}{2014}).

\bibitem{LoperJPhysChem2014}
\bibinfo{author}{L\"{o}per, P.} \emph{et~al.}
\newblock \bibinfo{title}{Complex Refractive Index Spectra of
  CH$_3$NH$_3$PbI$_3$ Perovskite Thin Films Determined by Spectroscopic
  Ellipsometry and Spectrophotometry}.
\newblock \emph{\bibinfo{journal}{J. Phys. Chem. Lett.}}
  \textbf{\bibinfo{volume}{6}}, \bibinfo{pages}{66--71} (\bibinfo{year}{2014}).

\bibitem{ManserNatPhoton2014}
\bibinfo{author}{Manser, J.~S.} \& \bibinfo{author}{Kamat, P.~V.}
\newblock \bibinfo{title}{Band filling with free charge carriers in organometal
  halide perovskites}.
\newblock \emph{\bibinfo{journal}{Nat. Photon.}} \textbf{\bibinfo{volume}{8}},
  \bibinfo{pages}{737--743} (\bibinfo{year}{2014}).

\bibitem{WuJACS2015}
\bibinfo{author}{Wu, X.} \emph{et~al.}
\newblock \bibinfo{title}{Trap States in Lead Iodide Perovskites}.
\newblock \emph{\bibinfo{journal}{J. Am. Chem. Soc.}}
  \textbf{\bibinfo{volume}{137}}, \bibinfo{pages}{2089--2096}
  (\bibinfo{year}{2015}).

\bibitem{KaindlNature2003}
\bibinfo{author}{Kaindl, R.~A.}, \bibinfo{author}{Carnahan, M.~A.},
  \bibinfo{author}{Hagele, D.}, \bibinfo{author}{Lovenich, R.} \&
  \bibinfo{author}{Chemla, D.~S.}
\newblock \bibinfo{title}{Ultrafast terahertz probes of transient conducting
  and insulating phases in an electron-hole gas}.
\newblock \emph{\bibinfo{journal}{Nature}} \textbf{\bibinfo{volume}{423}},
  \bibinfo{pages}{734--738} (\bibinfo{year}{2003}).

\end{thebibliography}

\begin{addendum}
 \item DC acknowledges funding from CFI, NSERC, and FQRNT and DV acknowledges a scholarship from the National Council of Science and Technology in Mexico (CONACYT). The Swedish Energy Agency (STEM), the Swedish Research Council, the Knut \& Alice Wallenberg foundation and the European Research Council (Advanced Investigator Grant to VS, 226136-VISCHEM) are acknowledged. The nanometer Consortium at Lund University (nmc@LU) is also acknowledged. At Northwestern University the work was supported by the grant SC0012541 funded by the U.S. Department of Energy Office of Science.
 \item[Competing Interests] The authors declare that they have no
competing financial interests.
 \item[Correspondence] Correspondence and requests for materials
should be addressed to D.G.C. \\(email: cooke@physics.mcgill.ca) and V. S. (email: Villy.Sundstrom@chemphys.lu.se).
\end{addendum}

\end{document}